\documentclass[prb,floatfix,twocolumn,showpacs,amsmath,amssymb]{revtex4}
\usepackage{graphicx}
\usepackage{dcolumn}
\usepackage{bm}

\begin{document}
\title{Unconventional metal-insulator transition in two dimensions}

\author{Manuela Capello, Federico Becca, Seiji Yunoki, and Sandro Sorella}
\affiliation{
CNR-INFM Democritos National Simulation Center and International School for 
Advanced Studies (SISSA), via Beirut 2-4, I-34014 Trieste, Italy
}

\date{\today}

\begin{abstract}
We show, by using a correlated Jastrow wave function and a mapping
onto a classical model, that the two-dimensional Mott transition in a simple
half-filled one-band model can be unconventional and very similar to the
binding-unbinding Kosterlitz-Thouless transition of vortices and anti-vortices, 
here identified by empty and doubly occupied sites.
Within this framework, electrons strongly interact with collective plasmon
excitations that induce anomalous critical properties on both sides of the
transition. In particular, the insulating phase is characterized by
a singular power-law behavior in the photoemission spectrum, that can be
continuously connected to the fully-projected insulating state, relevant to 
strongly correlated low-energy models.
\end{abstract}


\maketitle

\section{introduction}\label{intro}

The metal-insulator transition (MIT) driven by electron interaction, the
so-called Mott transition,~\cite{mott} is one of the most challenging issues 
in modern solid state physics, especially because of its possible connections 
with other phenomena, like high-temperature superconductivity.
The prototypical model to study the MIT is the one-band Hubbard model, 
where the transition shows up if the ratio between the bandwidth 
$W$ and the on-site Coulomb repulsion $U$ is varied.~\cite{imada}
In a pioneering work, Brinkman and Rice~\cite{rice} argued that the MIT could 
be continuous, however, their approach led to a Mott insulator where charge 
fluctuations are completely frozen. Recent developments, based on dynamical 
mean-field theory calculations,~\cite{dmft} are able to give a more realistic 
description of the MIT where, at zero temperature, a wide gap in the 
excitation spectrum opens immediately in the insulator.
However, within this approach, the role of the dimensionality is not taken into
account, whereas its relevance comes out from recent experiments on
Si MOSFET's~\cite{kravchenko} and organic materials.~\cite{kanoda}
Moreover, the situation can be radically different whenever the Hamiltonian 
contains a true long-range Coulomb interaction, as pointed out in the 
original Mott argument.~\cite{mott}

In this work, we focus our attention on a correlated 
wave function (WF), which is expected to closely describe the MIT in two 
dimensional (2D) systems with long-range Coulomb interactions.
The important fact is that, in the strong-coupling regime, we can 
adiabatically connect our WF with the fully-projected one, usually
considered to describe systems in the limit of infinite Coulomb repulsion,
e.g., the so-called resonating valence-bond 
state.~\cite{anderson,randeria,lee,dagotto}
Moreover, the metallic phase has no quasiparticles defined, showing low-energy
properties similar to the one-dimensional Luttinger liquid.

The paper is organized as follow: In Sec.~\ref{mapping} we show the mapping
between the quantum wave function and a classical model at finite temperature
and in Sec.~\ref{results} we show our numerical results and we draw the 
conclusions.

\section{Classical Mapping}\label{mapping}

Let us discuss how to construct a WF for correlated insulators. 
In general, starting from the ground state $|\Psi_0 \rangle$ of a 
system with $N$ electrons, with energy $E_0$, it is possible 
to construct simple variational states for the lowest-energy excitations. 
For instance, in analogy with the Feynman's construction
for the liquid Helium,~\cite{feynman} the plasmon excitation with momentum $q$ 
is given by:
\begin{equation}
|\Psi_q \rangle = n_q |\Psi_0 \rangle,
\end{equation}
where $n_q$ is the Fourier transform of the local electron density. 
Its variational energy is
\begin{equation}
E_q = E_0 + \frac{\langle -k \rangle q^2}{2N_q},
\end{equation}
where $\langle k \rangle$ is the ground-state kinetic energy per particle and 
\begin{equation}
N_q= \frac{\langle \Psi_0| n_{q} n_{-q} |\Psi_0 \rangle}
{\langle \Psi_0|\Psi_0 \rangle} 
\end{equation}
is its static charge structure factor. When applied to an insulator, 
with gapped charge excitations, this implies that, for small momenta,
$N_q \sim q^2$.
This argument has very general consequences on the form of $|\Psi_0 \rangle$, 
that do not depend on the particular microscopic model. To this purpose,
let us denote an electronic configuration by the positions $\{ x \}$ of
the particles. For all the operators $\theta$ that depend only on such
positions, e.g., the structure factor itself, the quantum average 
\begin{equation}
\langle \theta \rangle = \frac{\langle \Psi_0 | \theta |\Psi_0 \rangle}
{\langle \Psi_0|\Psi_0 \rangle}
\end{equation}
can be written in terms of the {\it classical} distribution
$|\Psi_0(x)|^2 = |\langle x|\Psi_0 \rangle|^2 / 
\sum_{x^\prime} |\langle x^\prime|\Psi_0 \rangle|^2$, as
\begin{equation}
\langle \theta \rangle = \sum_x \langle x|\theta|x \rangle |\Psi_0(x)|^2.
\end{equation}
Since $|\Psi_0(x)|^2$ is a positive quantity, we can define an appropriate
correspondence between the WF and an effective potential $V(x)$:
\begin{equation}\label{classical}
|\Psi_0(x)|^2 = e^{-V(x)}.
\end{equation}
The size consistency of the WF implies that the potential 
$V(x)$ is extensive, namely of order $N$ for typical configurations.
In the limit of strong Coulomb interactions, there are small charge 
fluctuations and, therefore, we can safely assume that only the two-body 
term is relevant and all multi-particle interactions are negligible. 
This leads to the quadratic potential
\begin{equation}
V(x) = \sum_{q \ne 0} v_q^{eff} n_{q}(x) n_{-q}(x),
\end{equation}
being $n_{q}(x)$ the Fourier transform of the local density of the 
configuration $|x \rangle$. To obtain the expected behavior 
\begin{equation}
N_q = \sum_x n_{q}(x) n_{-q}(x) e^{-V(x)} \sim q^2,
\end{equation}
the effective potential must diverge as  
\begin{equation}
v_q^{eff} = \frac{ \pi}{T^{eff} q^2} + \mbox{less singular terms}.
\end{equation}
Here $T^{eff}$ can be considered as the effective temperature of 
classical charges interacting through a potential $\pi/q^2$.
Within this choice of $v_q^{eff}$, $N_q \sim q^2$ is generally valid 
and can be understood by considering
$n_q$ as a complex continuous variable, so that the classical average 
of $n_q n_{-q}$ turns into a standard Gaussian integral, yielding  
\begin{equation}
N_q \sim \frac{1}{v_q^{eff}} = \frac{T^{eff} q^2}{\pi}. 
\end{equation}
It should be noted that the fully-projected 
wave function with no charge fluctuations, and therefore $N_q=0$ for 
$|q| \ne 0$, is recovered when $T^{eff} \to 0$.

\section{Results}\label{results}

Let us now consider a general one-band fermionic system in 2D, 
in which every site of a
square lattice can be either empty, singly occupied, being 
the electron either with spin up or down, or doubly occupied.
A true Mott insulator, that does not break any lattice symmetry,
cannot be represented by a simple WF containing a single determinant, and,
at this stage, it is useful to define a state that is simple enough
and yet is compatible with the predicted form of $v_q^{eff}$ in the 
insulating phase. A straightforward way to modify the effective potential 
determined by an uncorrelated determinant $|{\cal D} \rangle$ is obtained by 
taking into account an appropriate Jastrow factor ${\cal J}$:
\begin{equation}\label{wf}
|\Psi \rangle = {\cal J} |{\cal D} \rangle,
\end{equation}
here $|{\cal D} \rangle$ is an electronic determinant that will be specified 
in the following and ${\cal J}$ is a Jastrow term that depends upon the 
electronic density:
\begin{equation}
{\cal J} = \exp \left[ -\frac{1}{2} \sum_{q \ne 0} v_q n_{q} n_{-q} \right],
\end{equation}
where $v_q$ is the Jastrow potential, whose small-$q$ behavior is given by:
\begin{equation}\label{veff}
v_q = \frac{\pi \beta}{2\, [2-(\cos q_x +\cos q_y)]} \sim 
\frac{ \pi \beta}{q^2},
\end{equation}
$\beta$ fixing its strength.

At half filling, in one-dimensional electronic systems, 
we found~\cite{capello,capello2} that the singular Jastrow 
$v_q \sim \pi \beta/q^2$ always leads to an insulator, for any positive $\beta$.
In 2D the situation is different and a much more interesting 
scenario is obtained, with a phase transition as a function of the 
correlation strength $\beta$. Indeed, given the behavior of the Jastrow
$v_q\sim \pi \beta/q^2$, the potential $V(x)$ of 
Eq.~(\ref{classical}) turns out to be the one of the classical Coulomb 
gas model (CGM). 
In this approach, particles with charge $q_i$, corresponding to empty 
($q_i=1$) and doubly ($q_i=-1$) occupied sites, interact through a Coulomb 
potential in a neutral background, represented by singly 
occupied sites ($q_i=0$). In the half-filled case, 
there is an equal number of empty and doubly occupied sites, implying the 
charge neutrality of the CGM. 
The fugacity $z$ of the CGM, that sets the average value of the 
charges, can be identified with the on-site Gutzwiller term in the Jastrow 
potential, i.e., $z={\rm exp}(-g)$, where $g$ is the Gutzwiller parameter.

\begin{figure}
\includegraphics[width=0.43\textwidth]{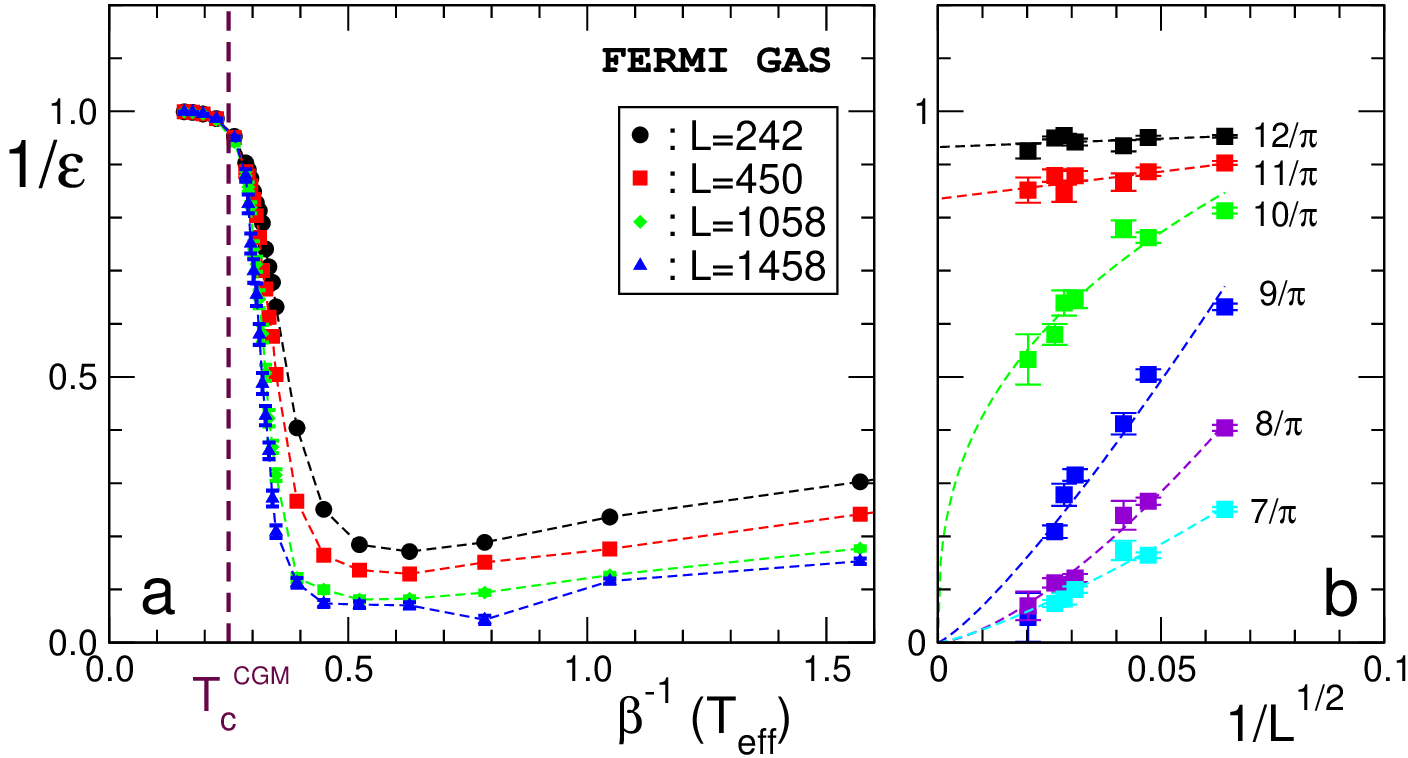}
\caption{\label{fig:jumpa}
(Color online)
Inverse of the dielectric function $1/\epsilon$ [see Eq.~(\ref{epsilon})] 
for the free-electron determinant.
Left panel: $1/\epsilon$ as a function of the effective temperature $1/\beta$
and for different sizes $L$ of the cluster. 
The critical temperature of the classical Coulomb gas model $T^{CGM}_c$ 
is marked with a dashed line for a comparison.
Right panel: size scaling of $1/\epsilon$ for various $\beta$. 
}
\end{figure}

\begin{figure}
\includegraphics[width=0.43\textwidth]{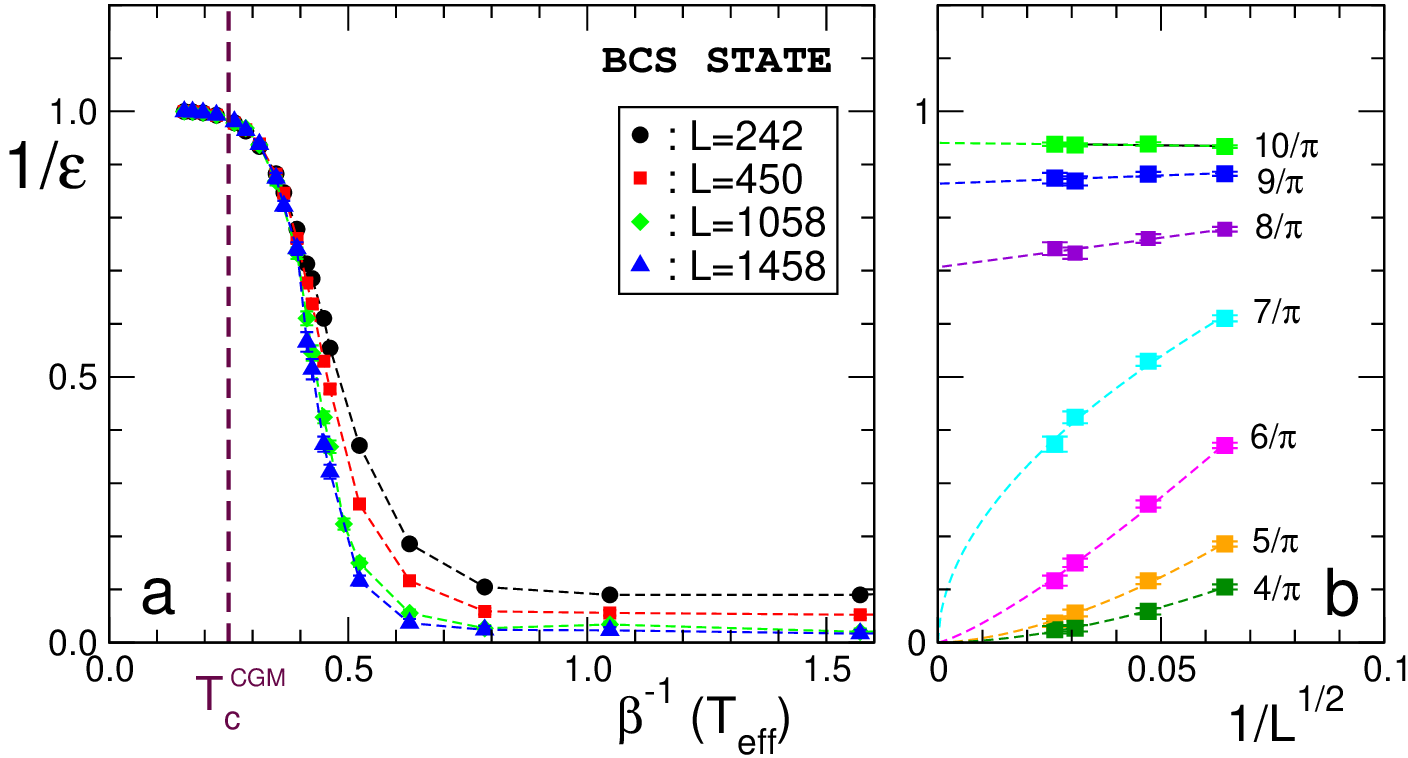}
\caption{\label{fig:jumpb}
(Color online)
The same as in Fig.~\ref{fig:jumpa} for a gapless BCS state with 
$\Delta/t=1.1$ and $d_{x^2-y^2}$ symmetry.
}
\end{figure}

The classical CGM in 2D is particularly interesting because 
it shows a Kosterlitz-Thouless (KT) transition at a finite 
temperature $T_c^{CGM}$.~\cite{minnhagen}
This transition is transparent from the classical dielectric 
function:
\begin{equation}\label{epsilon}
\frac{1}{\epsilon} = \lim\limits_{q\to 0} \left[ 1 - \frac{2\pi}{T^{eff} q^2} N_q
\right],
\end{equation}
where $T^{eff}$ is the temperature of the classical model.
The charge structure factor is quadratic at small momenta, 
i.e., $N_q \sim \alpha q^2$, for all temperatures,
but the coefficient $\alpha$ changes discontinuously at $T_c^{CGM}$.
Above $T_c^{CGM}$, the CGM is in the plasma phase, i.e., a metallic phase with 
infinite dielectric function, perfect screening, and exponential correlation 
functions.  
On the other hand, below $T_c^{CGM}$, the CGM is in the confined phase, 
with a finite dielectric constant. In this phase the charges are bound 
together forming dipoles, that, because of their residual interaction, 
induce power-law correlations. At the transition, the inverse of the 
dielectric function has a finite jump, changing from zero, in the plasma 
phase, to a finite value, in the confined phase.

A similar mapping between a quantum state and a 
classical model has been emphasized also in the context of the fractional 
quantum Hall effect: the Laughlin WF can be related to a classical 
system with particles interacting through a logarithmic 
potential.~\cite{laughlin}
However, in this case, all the particles have the same charge, forming a
one-component plasma, and by varying the strength of the potential, 
there is a first-order transition between an incompressible fluid and a 
Wigner crystal.~\cite{levesque}
The peculiarity of our approach is that, due to the mapping onto the 
two-component CGM, it is possible to connect continuously the plasma phase 
to the insulating one.

\begin{figure}
\includegraphics[width=0.40\textwidth]{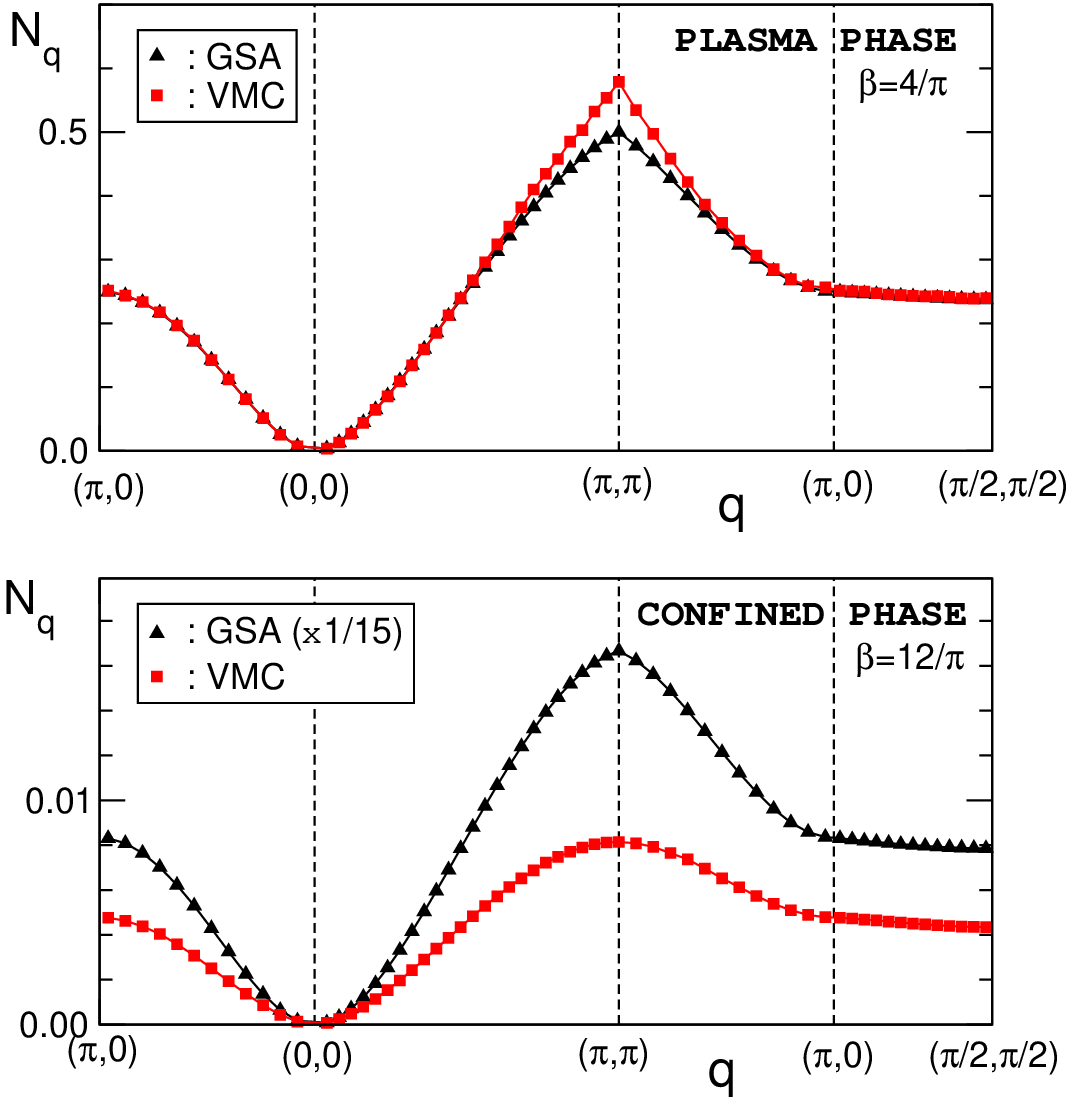}
\caption{\label{fig:rpa}
(Color online)
Equal-time density structure factor $N_q$ for the correlated wave function 
of Eq.~(\ref{wf}) (full squares), compared to the same quantity calculated 
within the Gaussian approximation [indicated as GSA and given 
by Eq.~(\ref{enneq})], (full triangles) for $\beta=4/\pi$ 
(upper panel) and $\beta=12/\pi$ (lower panel, notice the 
different scale of the GSA data). 
}
\end{figure}

In the following, we show that, in analogy with the classical CGM, 
also in the case of fermionic systems at zero temperature, a KT-like 
transition is found by varying the correlation strength that can be tuned by 
the Jastrow strength $\beta$. However, the existence of the fermionic part 
induces non-trivial properties for the two phases involved, that are not 
present in the classical problem.
For example the uncorrelated part of the WF may contribute to the 
expression of the effective temperature $T^{eff}$, as shown below.
Whenever the square of the WF describes the plasma phase of the corresponding 
classical model, we can safely assume that the Gaussian fluctuations
are exact for small $q$'s and the classical temperature can be determined by 
imposing $1/\epsilon=0$ in Eq.~(\ref{epsilon}), 
namely $T^{eff}= 2 \pi \lim_{q \to 0} N_q/q^2$.
In the language of quantum states, the Gaussian approximation leads
to the well-known expression
\begin{equation}\label{enneq}
N_q = \frac{N^0_q}{1 + 2 v_q N^0_q},
\end{equation}
where $N^0_q$ is the charge structure factor of the uncorrelated 
determinant $|{\cal D}\rangle$.~\cite{gaskell,reatto}
The previous form of $N_q$ allows us to identify 
the effective temperature as: 
\begin{equation}
\label{Teff}
\frac{1}{T^{eff}}=\beta+\frac{\alpha_0}{2 \pi}
\end{equation}
where 
$\alpha_0=\lim_{q \to 0} q^2/N^0_q$.

In order to show the general validity of our approach, we consider the case 
of a free-electron determinant, obtained by occupying the lowest-energy states 
in the tight-binding model with dispersion $E_k = -2t (\cos k_x +\cos k_y)$ 
and a gapless BCS state with a superconducting order parameter 
$\Delta_k=\Delta(\cos k_x -\cos k_y)$.
In these cases $\alpha_0=0$ and, therefore, the effective temperature
in Eq.~(\ref{Teff}) is determined only by the Jastrow coefficient, 
namely $T^{eff}=1/\beta$. 
In Fig.~\ref{fig:jumpa}, we report the inverse of the dielectric function
for the free-electron determinant and different sizes $L$ of the system at 
half filling, i.e., $N=L$. 
In order to have closed-shell states for $|{\cal D} \rangle$, we used 2D 
square lattices tilted by $45^\circ$ (i.e., with $L=2 l_x^2$ and $l_x$ odd) 
and periodic boundary conditions. 
By increasing $L$, the curves show
a steeper and steeper shape in the vicinity of the critical temperature $T_c$.
This result is further confirmed by the size scaling of $1/\epsilon$, which 
clearly supports the existence of a finite jump in the thermodynamic limit:
$1/\epsilon \to 0$ for $T^{eff}>T_c$, whereas $1/\epsilon \to {\rm const}$ for 
$T^{eff}<T_c$. Interestingly, $T_c$ depends slightly upon the choice of
the uncorrelated determinant (see for comparison Fig.~\ref{fig:jumpb} for the 
gapless BCS state)
and is quite close to the CGM critical temperature $T_c^{CGM}=1/4$.
These results give an important and transparent insight into the 
strong-coupling limit described by the fully-projected WF,~\cite{anderson}
that can be connected to our WF by letting $\beta \to \infty$, i.e.,
$T^{eff} \to 0$.
Indeed, in the confined phase for $T^{eff}<T_c$, the classical KT scaling 
equations of the CGM flow to fixed points with zero fugacity: this translates 
into the fact that the fully-projected state represents the fixed-point of the 
correlated WFs describing the 2D Mott insulating phase. 
Therefore, in the confined regime, the ground-state properties 
are universal and represented by the ones of the fully-projected WF.
In this respect, the total projection is not an unrealistic assumption and can 
accurately reproduce the low-energy physical properties of a strongly 
correlated system.
On the other hand, for $T^{eff}>T_c$ the classical KT scaling equations flow to 
strong coupling and are useful only close to the transition point.

\begin{figure}
\includegraphics[width=0.40\textwidth]{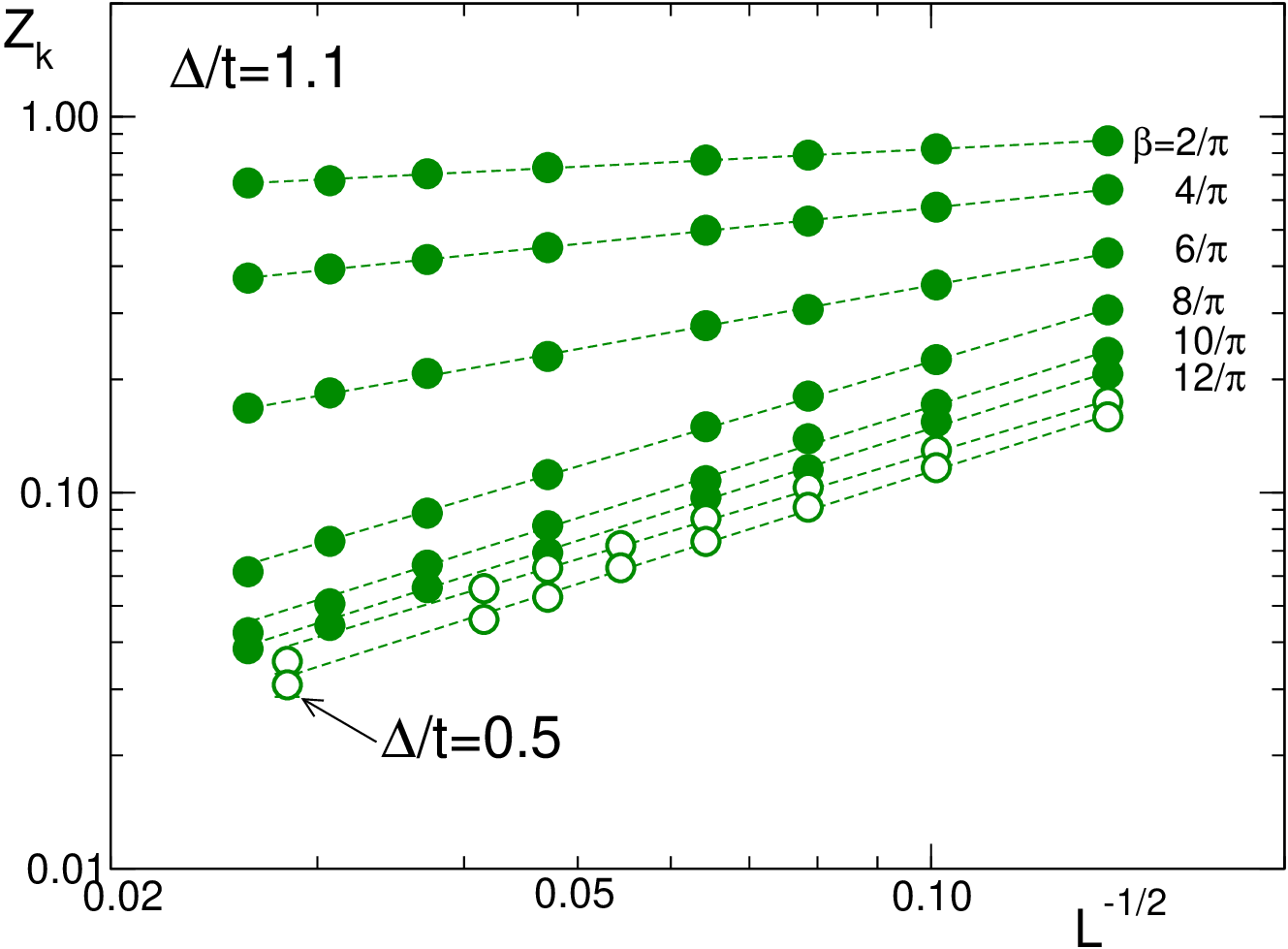}
\caption{\label{fig:zeta}
(Color online)
Quasiparticle weight $Z_k$ at $k=(\pi/2,\pi/2)$ for the gapless BCS state 
with $\Delta/t=1.1$ and $d_{x^2-y^2}$ symmetry as a function of $L$
and for different Jastrow strengths $\beta$ (full circles). 
The case of the fully-projected wave function (empty circles) is also 
reported for $\Delta/t=1.1$ and $0.5$.
}
\end{figure}

\begin{figure}
\includegraphics[width=0.40\textwidth]{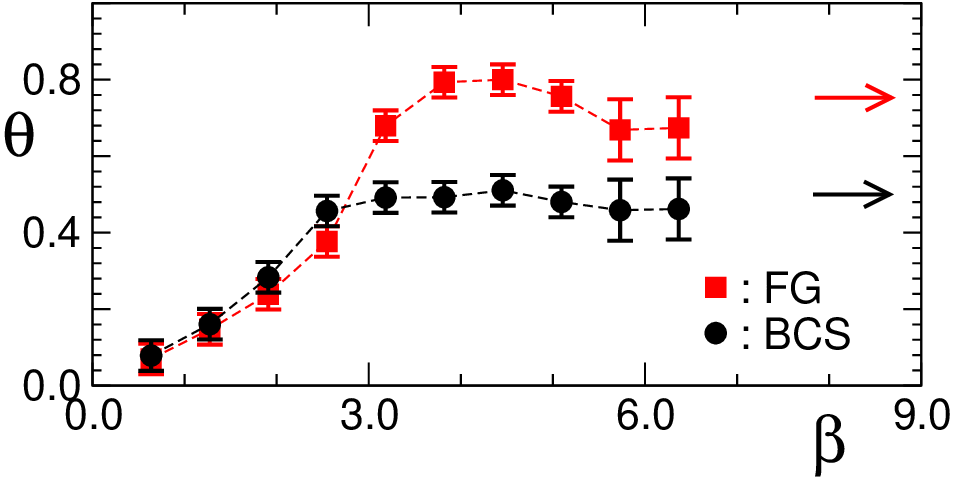}
\caption{\label{fig:theta}
(Color online)
The behavior of $\theta$ [the exponent of the quasiparticle weight, see
Eq.~(\ref{powerlaw})] as a 
function of $\beta$ for BCS state (full circles) and the Fermi gas (FG) 
determinant (full squares).
The value of the fully-projected states are also reported (arrows).
}
\end{figure}

In Fig.~\ref{fig:rpa} we show that in the plasma phase the Gaussian 
approximation, given by Eq.~(\ref{enneq}), is very accurate, not only for 
small $q$'s (where it is exact) 
but also for large momenta.  In this case, the cusp singularity in $N_q$ 
for $Q=(\pi,\pi)$, related to the Friedel oscillations, is not removed, 
even though $N_q \sim q^2$ at low momenta. Thus, for $T^{eff}>T_c$, 
the WF~(\ref{wf}) describes a ``Coulomb metal'', with $N_q \sim q^2$ at 
small $q$'s but with the sign of the Fermi surface at large momenta.
As shown previously in the limit of infinite Coulomb 
repulsion~\cite{valenti} or in the 
low-density regime,~\cite{bares} this WF has low-energy properties 
similar to one-dimensional Luttinger liquid conductors, where the absence 
of a jump in the momentum distribution is replaced by a weaker singularity, 
yielding to $2 k_F$ and $4 k_F$ power-law density correlations.
It is important to emphasize that, in the quantum case, the power-law 
correlations come from the large momentum singularity, that are absent in 
the classical CGM.~\cite{minnhagen}
Indeed, in the quantum state, the subleading corrections in the
classical potential of Eq.~(\ref{classical}) are very important 
and can actually turn the CGM exponential correlations to power laws 
in the plasma phase, and {\it vice-versa} in the confined phase.
On the other hand, in the confined phase the Gaussian approximation
is not adequate both at small and large momenta, see 
Fig.~\ref{fig:rpa}. Indeed, at small $q$'s, 
the coefficient of the quadratic term is not simply given by the Gaussian
approximation and, more importantly, the strong Jastrow factor washes out 
completely the singularities of $N^0_q$, leading to a smooth charge-structure 
factor, a genuine fingerprint of an insulating phase.

\begin{figure}
\includegraphics[width=0.40\textwidth]{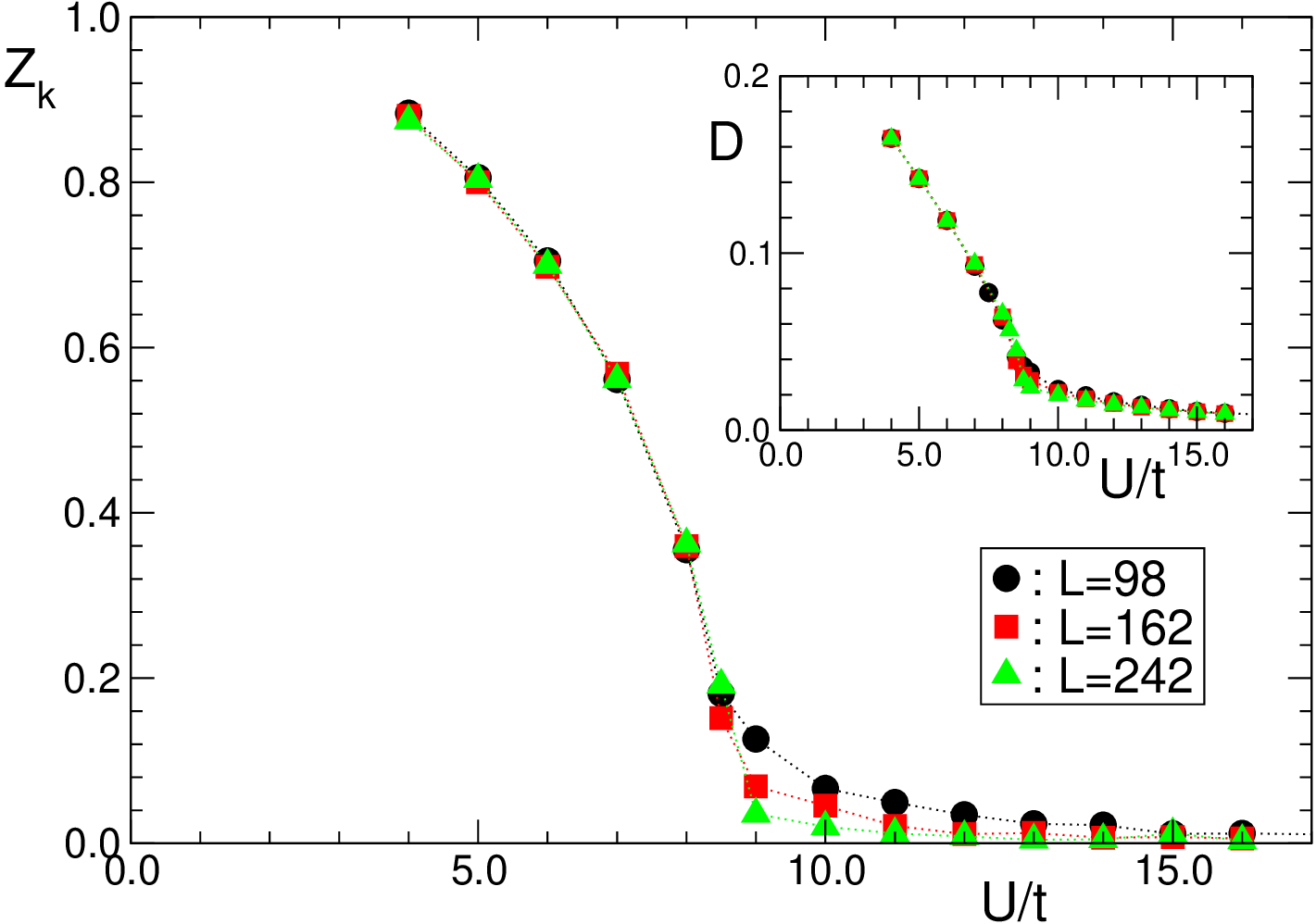}
\caption{\label{fig:transition}
(Color online)
Quasiparticle weight $Z_k$ at $k=(\pi/2,\pi/2)$ of the optimized  
{\it paramagnetic} wave function containing a Jastrow factor 
applied to the Fermi gas as a function of the interaction $U/t$ 
in the Hubbard model, for three different sizes of 
the system. Inset: the number of double occupancies $D$ as a function of $U/t$.
}
\end{figure}

\begin{figure}
\includegraphics[width=0.40\textwidth]{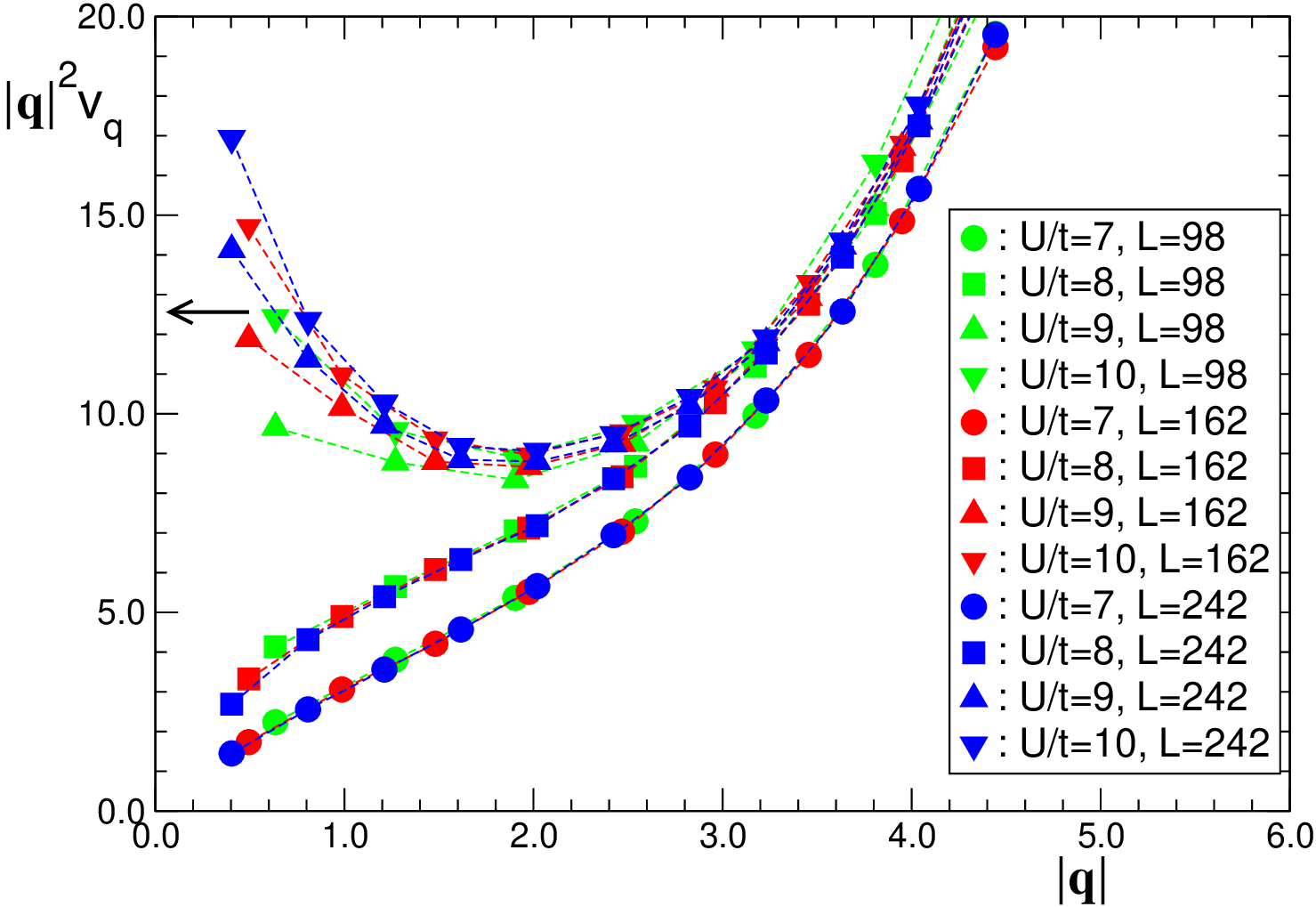}
\caption{\label{fig:vq}
(Color online)
Optimized Jastrow potential $v_q$, multiplied by $|q|^2$, for the Hubbard model
as a function of $|q|$ [in the (1,1) direction] for different sizes of the 
cluster and ratios $U/t$. The arrow indicates $\pi/T_c^{CGM}$, the 
expected value of $\lim_{q \to 0} v_q |q|^2$ at the classical transition point.
}
\end{figure}

In order to further characterize the two phases, we consider the 
quasiparticle weight
\begin{equation}
Z_k = \frac{|\langle \Psi_{N-1}|  c_{k,\sigma}  |\Psi_N \rangle|^2}
{\langle \Psi_N| \Psi_N \rangle \langle \Psi_{N-1}| \Psi_{N-1} \rangle},
\end{equation}
where $|\Psi_N \rangle$ and $|\Psi_{N-1} \rangle$ are the WF 
with $N$ and $(N-1)$ particles, $c_{k,\sigma}$ is the destruction operator of 
a particle of momentum $k$ and spin $\sigma$. 
In particular, the wave function with $N-1$ particles is constructed from
$|\Psi_N \rangle$:
\begin{equation}
|\Psi_{N-1} \rangle = {\cal J} c_{k,\sigma} |{\cal D} \rangle.
\end{equation}
In a previous work,~\cite{bares} it was argued that the singular Jastrow
factor can induce non-Fermi liquid properties, and in particular a 
vanishing $Z_k$ at the Fermi surface. In Fig.~\ref{fig:zeta}, we report 
$Z_k$ for $k=(\pi/2,\pi/2)$ and for different Jastrow strengths $\beta$.
We find that the quasiparticle weight vanishes with a power-law behavior:
\begin{equation}\label{powerlaw}
Z_k \sim L^{-\theta}
\end{equation}
both in the confined and in the plasma phase, 
with an exponent $\theta$ that depends upon $\beta$ and the type of the 
uncorrelated state. In the plasma phase, $\theta$ varies continuously with 
the Jastrow strength $\beta$ and there is no appreciable dependence on the 
uncorrelated determinant. On the other hand, in the confined phase, the 
exponent is constant, i.e., $\theta \simeq 1/2$ for the BCS state and 
$\theta \simeq 3/4$ for the free-electron state, and equal to the value
found for the fully projected WF, as shown in Fig.~\ref{fig:theta}. 
It must be mentioned that, for the BCS state, $\theta$ does not depend upon 
the value of the superconducting order parameter $\Delta$ 
(see Fig.~\ref{fig:zeta}), indicating the universal properties of the 
confined phase.

Our results show that it is possible to describe a continuous MIT in 2D 
electronic systems with a Jastrow correlated WF. We characterized both the
metallic region, with a zero-quasiparticle weight $Z_k=0$, and the insulator,
which can be continuously connected to the totally projected Gutzwiller 
WF. Of course, other scenarios are possible for the MIT, e.g., the one
proposed in the infinite-dimensional limit.
Indeed, whenever the metallic phase has $N_q \sim c |q| + d q^2$ with $c>0$, 
the MIT is not described by the functional form~(\ref{veff}) of the Jastrow:
By approaching the transition 
from the metallic phase, we enter directly into the confined phase with 
a quadratic charge structure factor at small momenta, 
i.e., $c \to 0$ with a large finite $d$ at the critical point.
In this case, in the metallic region, a less singular Jastrow factor 
$v_q \sim 1/|q|$ is expected, leading to a finite quasiparticle weight.

Finally, we would like to comment about the possibility to stabilize the
``Coulomb metal'' phase in a microscopic model. 
For simplicity, let us consider the one-band Hubbard model on the square 
lattice with nearest-neighbor hopping:
\begin{equation}
H= -t \sum_{\langle i,j \rangle, \sigma} c^\dag_{i,\sigma} c_{j,\sigma} + h.c.
+U \sum_i  n_{i,\uparrow} n_{i,\downarrow},
\end{equation}
where $c^{\dag}_{i,\sigma}$ creates an electron with spin $\sigma$ at the
site $i$, $n_{i,\sigma}= c^{\dag}_{i,\sigma} c_{i,\sigma}$ is the density
operator at the site $i$. In the following, we will consider 
a {\it paramagnetic} state, by taking a projected Fermi gas wave function,
as described above, and by minimizing the variational energy for the 
determination of the Jastrow factor ${\cal J}$.
In particular, by using the method described in Ref.~\onlinecite{sorella}, 
we are able to optimize all the independent Jastrow parameters in the real 
space $v_{i,j}$ (i.e., the Fourier transform of $v_q$).
Within this approach, that neglects magnetic phases, we obtain a MIT
for $U_c/t = 8.5 \pm 0.5$. In the weak coupling regime, for $U<U_c$, 
we obtain a Fermi liquid with a finite quasiparticle weight, whereas at strong
couplings, i.e., for $U>U_c$, we have an insulating phase with a vanishing 
$Z_k$, see Fig.~\ref{fig:transition}.~\cite{notez}
Moreover, the calculation of the double occupancy $D$ clearly indicates that
the transition is continuous and the insulating phase still possess finite
charge fluctuations, see inset of Fig.~\ref{fig:transition}.
As discussed above, in the metallic region, we find that $v_q \sim 1/|q|$
(see Fig.~\ref{fig:vq}). Unfortunately, as soon as we enter in the insulating 
phase, $\lim_{q \to 0} v_q \; |q|^2$ defines an effective $\beta$ which is 
larger than the critical value for the KT transition, and, therefore, no 
evidence for the ``Coulomb metal'' is found.~\cite{note}
Indeed, we expect that the optimized Jastrow factor $v_q$, 
containing subleading corrections with respect to Eq.~(\ref{veff}), will define 
a critical $\beta$ very close to the value of the classical CGM, 
i.e., $\beta_c = 1/T_c^{CGM} = 4$.
Therefore, in light of the results of Fig.~\ref{fig:vq}, the stabilization
of the ``Coulomb metal'' seems to be very unlikely: Although there are large 
size effects around $U_c$, we have a clear evidence that 
$\lim_{q \to 0} v_q |q|^2 \to 0$ for $U<U_c$ and 
$\lim_{q \to 0} v_q |q|^2 \gtrsim 4 \pi$ for $U>U_c$ (see Fig.~\ref{fig:vq}).

On the other hand, we can safely predict the occurrence of the novel KT-like 
scenario described above in 2D systems with long-range (logarithmic) 
interaction. 
In this case, the application of the Gaussian approximation for small 
interaction and our ansatz for the insulating phase imply the presence of a 
transition of the type considered here. It is remarkable that the proposed 
picture crucially depends on the long-range nature of the Coulomb 
interaction, recalling Mott's original idea. 
In this regard, it should be mentioned that the original Mott 
argument for a discontinuous metal-insulator transition, driven by the 
long-range Coulomb interaction, cannot be applied in 2D. 
In such case there always exists a bound state for two opposite 
charges interacting with the screened Coulomb potential, so that, 
according to this argument, no metallic phase with unbound charges 
is possible. However, this is clearly an artifact of such mean-field argument, 
since for small interaction the random-phase approximation leads to an 
anomalous metallic state (see Ref.~\onlinecite{bares}).

We thank M. Fabrizio, A. Parola, and E. Tosatti for the valuable and fruitful 
interaction during the accomplishment of this project. 
We also thank C. Castellani, N. Nagaosa, T. Senthil, and X.-G. Wen 
for interesting discussions.  
This work has been supported by INFM and MIUR (COFIN 2004 and COFIN 2005).


\begin{thebibliography}{99}
\bibitem{notez} In order to minimize the size effects on $Z_k$, we
   calculate the momentum distribution $n_k$ and extract, by fitting around
   the known Fermi surface, the value of the jump.
\bibitem{note} This outcome is expected to be the same for generic short-range 
   Hamiltonians, and, therefore, the presence of a frustrating hopping term
   does not help to stabilize the ``Coulomb metal'' phase.
\end{thebibliography}
\end{document}